\documentclass[preprint, 12pt,nofootinbib]{revtex4-1}

\usepackage{graphicx}
\usepackage{cancel}
\usepackage{amssymb}
\usepackage{textcomp}
\usepackage{amsmath}
\usepackage{bm}
\usepackage{times}
\usepackage{epsfig}
\usepackage{color}

\def\beq{\begin{equation}}
\def\eeq{\end{equation}}
\def\be{\begin{equation}}
\def\ee{\end{equation}}
\def\bea{\begin{eqnarray}}
\def\eea{\end{eqnarray}}
\def\to{\rightarrow}

\begin{document}
\title{Can Up FCNC solve the $\Delta A_{CP}$ puzzle?}
\author{Kai Wang}
\email{wangkai1@zju.edu.cn}
\author{Guohuai Zhu}
\email{zhugh@zju.edu.cn}
\affiliation{Zhejiang Institute of Modern Physics and Department of Physics, Zhejiang University, Hangzhou, Zhejiang 310027, CHINA}

\begin{abstract}
We investigate the attempt using flavor violation gauge interaction in the up sector
to explain the LHCb recently observed large $\Delta A_{CP}$ ($A_{CP}(D^{0}\to K^{+}K^{-})-A_{CP}(D^{0}\to \pi^{+}\pi^{-})$).  We study an Abelian model that only right-handed up quarks is charged under it and the $1-3$ coupling is maximized. The simultaneous $1-3$ $2-3$ mixing is realized by a quark mixing of $1-2$ generation.  Given the easy identification of top quark, the model can be directly tested by $\Delta F=1$ and $\Delta F=2$ processes at the hadron colliders as associated top production $g c \to t Z^{\prime}$ or same-sign top scattering $u u\to t t$. The direct search bounds are still consistent with the assumption that $ut$ and $ct$ couplings are equal but the same-sign top scattering bound is expected to be reached very soon. However, since there is no CKM-like suppression, the corresponding parameter space for generating $\Delta A_{CP}$ is completely excluded by the $D^{0}-\bar{D}^{0}$ mixing. We conclude that the up FCNC type models cannot explain the $\Delta A_{CP}$ while to be consistent with the $D^{0}-\bar{D}^{0}$ mixing constraint at the same time. On the other hand,  a model as SM with fourth family extension has better chance to explain the large $\Delta A_{CP}$ consistently.

\end{abstract}

\maketitle

\section{Introduction}
CP violation in the $D$ meson decay processes $c\to u q\bar{q}$ is highly suppressed in the standard model (SM).
Given its small SM expectation, CP violation in $D$ meson decay play important role to probe models beyond SM (see, for example \cite{kagan} and 
references therein). Recently, the LHCb collaboration has reported a measurement of difference in CP asymmetry, $A_{CP}(D^{0}\to K^{+}K^{-})-A_{CP}(D^{0}\to \pi^{+}\pi^{-})$ based on the data of 580~pb$^{-1}$ \cite{talk}. The measured difference in CP asymmetry,
\beq
\Delta A_{CP} =[-0.82\pm0.21(\text{stat.})\pm0.11(\text{sys.})]\%
\eeq
which corresponds to the SM prediction of $10^{-4}$. The deviation from the SM prediction is then 3.5~$\sigma$ evidence.
The CPV in $c\to u q\bar{q}$ arises from the interference between tree level amplitude of SM charged current and the QCD penguin amplitude. As a result of GIM-mechanism \cite{gim}, the QCD penguin amplitude completely vanish at the limit when internal quarks in penguins are massless. Within the SM framework, non-zero contribution to direct CP violation comes only 
from the bottom quark which is proportional to $V_{cb}^\ast V_{ub} m^{2}_{b}/m^{2}_{W}$. The CKM factor here is very small, suppressed by  $\lambda^{5}$. With additional loop factor suppression, the CP violation in $c\to u q\bar{q}$ is typically of ${\cal O}(10^{-4})$ in short distance calculation. It is unlikely that nonpertubative effects may enhance the direct CP violation to be above $10^{-3}$. 

The recent measure of difference in $D^{0}\to K^{+}K^{-}$ and  $D^{0}\to \pi^{+}\pi^{-}$ basically minimize the effect from indirect CP violation, namely CP violation in the $D^{0}-\bar{D}^{0}$ mixing. The significant $\Delta
 A_{CP}$ clearly indicates that the observed CP violation should occur in the $c\to u q\bar{q}$ decay directly.
To solve the anomaly, new physics is required to enhance the CP violation in $c\to u q\bar{q}$ decay \cite{perez, kagan2}.
One simple extension of SM is to introduce a fourth family of quarks and leptons, the fourth generation down quark $b^{\prime}$ of 400~GeV can enhance the penguin amplitude by ${\cal O}(10^{4})$ in mass squared but with suppression from quark mixing. The Cabibbo-Koboyashi-Maskawa quark mixing matrix (CKM) of fourth family is constrained by precision electroweak tests, for instance $\rho$-parameter and the $34$-mixing
of $\lambda$ is allowed \cite{ckm4}. We estimate the suppression from quark mixing of CKM4 is about $\lambda^{7}$ and the penguin amplitude is then enhanced by ${\cal O}(10^{2})$ from the fourth generation $b^{\prime}$ contribution. It does provide possible parameter space to accommodate the about $1\%$ CP violation. At the same time, the contribution to the $D^{0}-\bar{D}^{0}$ mixing is under control with additional CKM suppression. However, in this paper, we would like to focus on the other possibility involving top quark.

Top quark contribution is GIM violation.  As the heaviest known particle
that gets its mass via electroweak symmetry breaking (EWSB), top quarks couples to
the longitudinal polarized $W_{L}$ state strongly and leads to the enhancement as $m^{2}_{t}/m^{2}_{W}$ in the SM penguin of $b\to d_{i}$ transition.  On the other hand, in many new physics models, top quark very often appears in $c\to u$ transition and the $\Delta C=1$ decay like $c\to u q\bar{q}$ is very sensitive to such models.
More interestingly, since top quarks decay before hadronization and can be directly measured
at detectors, if the new physics involves top quarks, both $\Delta C=1$ and $\Delta C=2$ can also be  tested at the collider directly. $g c \to t Z^{\prime}$ and $c c \to t t$ (or $g u\to t Z^{\prime}$ and $uu \to tt$).

A recent anomaly observed at the Tevatron basically motivated most of such models of $c\to u$ transition involving the top quarks. CDF collaboration at Tevatron in this January reported the
reconstructed top quark forward-backward asymmetry in the semi-leptonic $t\bar{t}$ system.
The most significant deviation appears in the $t\bar{t}$ sample with large $t\bar{t}$ invariant mass $M_{t\bar{t}}$ while the others are mostly within 2~$\sigma$.
For $M_{t\bar{t}} > 450~\text{GeV}$, the forward-backward asymmetry for reconstructed top quark measured in the $t\bar{t}$ rest frame is
\beq
A^{t\bar{t}}_{FB}(M_{t\bar{t}} > 450~\text{GeV})= 0.475\pm 0.112~.
\eeq
The measurement corresponds to the SM prediction $A^{t\bar{t}}_{FB}(\textrm{CDF})=0.128$ \cite{hollik}
which includes both QCD ${\cal O}(\alpha^{3}_{s})$ and Electroweak ${\cal O}(\alpha^{2}_{s}\alpha)$ corrections. Again, this deviation appears as over 3~$\sigma$ \footnote{The CDF observation is not confirmed by the D$0$ collaboration. The unfolded D$\cancel{0}$ measurement of $A^{t\bar{t}}_{FB}(M_{t\bar{t}}>450~\text{GeV})=0.115\pm0.06$ which is within 1~$\sigma$ of SM prediction\cite{hollik}.  }. Since the other measurements in $t\bar{t}$ like total production rate $\sigma_{t\bar{t}}$
are in good agreement with the SM predictions, the proposals to solve the large $A_{FB}$ all
require destructive interference between the new physics and the SM $u\bar{u}\overset{g}\rightarrow t\bar{t}$, $d\bar{d}\overset{g}\rightarrow t\bar{t}$. In addition, since the anomaly corresponds to
a large $M_{t\bar{t}}$ region, the $t$-channel proposal
\cite{hitoshi, cheung, tim, james,jing} which maximize the asymmetry at Rutherford singularity $\theta=0$ match the basic feature of the measurement. Among the proposals, $t$-channel neutral current process interferes with the largest SM mode $u\bar{t}\overset{g}\rightarrow t\bar{t}$ and spin-correlation also maximize the positive forward-backward asymmetry \cite{hitoshi,james,jing}.  However, in order to explain the top quark forward-backward asymmetry puzzle, only significant $ut$ coupling is required and this is not sufficient to generate the $c\to u$ transition. In this paper, we investigate the possibility of generating $c\to u$ transition mediated by the top quark penguin.

Not surprisingly, the most stringent constraint would come from $D^{0}-\bar{D}^{0}$ mixing. Unlike the QCD penguin,  $D^{0}-\bar{D}^{0}$ mixing in the SM is dominated by the strange quark contribution. At amplitude level, the bottom quark contribution in the box diagram is suppressed by a factor of $\lambda^{10}$ while the corresponding strange quark contribution is only $\lambda^{2}$ suppression but the mass dependence is
still quadratic as $m^{2}_{q}/m^{2}_{W}$.

In the next section, we discuss the model setup. Then we study the
model parameter space required by the $\Delta A_{CP}$ measurement in the third section. In section IV, various constraints of the model
are discussed, in both low energy physics like $D^{0}-\overline{D}^{0}$ and collider experiments like same-sign top quark, inclusive $t\bar{t}$ search. We then conclude in the final section.

\section{Model}

In order to achieve large $c\to u$ transition
induced by top quark penguin, the new gauge interaction
must couple to both $\bar{t}c$ and $\bar{t} u$.
Flavor changing interactions in the SM can only be measured via electroweak charged current interactions. For the SM fermion rotation matrixes,  the left-handed ones get constrained from the CKM matrix $V_L^u (V_L^d)^\dagger = V_{CKM}$
but only the product instead of the $V^{u}_{L}$ and $V^{d}_{L}$ respectively. The rotations for the right-handed states are then completely unknown and this gives large degree of freedom.

We first study an Abelian model, a $U(1)_{X}$ gauge symmetry
under which only right-handed up-type quarks transform. With only
the SM particle contents, the $U(1)_{X}$ is anomalous so we expect
a UV completion theory. Presumably a much larger gauge group is broken
at very high energy and only a $U(1)_{X}$ survive to low energy
and is broken around TeV scale. In this paper, we don't discuss the
detail of the UV theory. Instead,
we concentrate on the low energy theory of the electroweak scale $U(1)_{X}$ gauge boson interacting
with the SM fermions.

In the flavor basis of $(u c t)$ , $U(1)_{X}$ is
\begin{eqnarray}
T=\lambda_{4}
=\begin{pmatrix}
0&0&1\\
0&0&0\\
1&0&0
\end{pmatrix}
\label{generator}
\end{eqnarray}
which shows only $ut$ couples to the $Z^{\prime}$.
As discussed earlier, there also exists degree of freedom
of right-handed up quark rotation.
We take a special choice of the rotation $V^{u}_{R}$ to illustrate the feature.
\begin{eqnarray}
V^{u}_{R}
=\begin{pmatrix}
\cos\theta&\sin\theta & 0\\
-\sin\theta&\cos\theta&0\\
0&0&1
\end{pmatrix}
\label{rotation}
\end{eqnarray}
The CP violating phase $e^{i\delta}$ can be easily included into the above rotation.

The effective lagrangian is then
\beq
{g_{X}\over \sqrt{2}}Z^{\prime}_{\mu} (\bar{t}_{R}\gamma^{\mu} u_{R} c_{\theta}+  \bar{t}_{R}\gamma^{\mu} c_{R} s_{\theta})
\eeq
where $c_{\theta}= \cos\theta$, $s_{\theta}= \sin\theta$, $g_{X}$ is
the coupling constant of the $U(1)_{X}$.

If the particle $Z^{\prime}$ is completely neutral which enable $Z^{\prime}$ to couple $\bar{t}u$, $\bar{u}t$ at the same time.
Both $u\bar{u}\to t\bar{t}$ and $uu\to t t$ exist and the first one dominates at the $p-\bar{p}$ collisions at Tevatron while the second one dominates at the Large Hadron Collider(LHC). Given its huge $u$-valence quark flux at the $p-p$ collider LHC, even with 7~TeV total energy, the $Z^{\prime}$ receives severe constrain from direct search of $uu\to tt$ \cite{jing}. To resolve the same-sign top puzzle, non-Abelian horizontal gauge symmetry models are proposed in \cite{hitoshi, james}.
In principle, a non-Abelian model where $(u_{R} t_{R})^{T}$ form a doublet under a $SU(2)_{X}$ and  can avoid large same-sign top quark production and box contribution to $D^{0}-\bar{D}^{0}$ mixing. However, the rotation to give $ct$ couplings
will generate large $uc$-mixing mediated by the $W^{3}$. $W^{^{\prime}\pm}$ and $W^{\prime 3}$ are nearly degenerate at the $SU(2)_{X}$ limit. Then the tree level $D^{0}-\bar{D}^{0}$ mixing is inevitable. The parameter space that generates the $\Delta A_{CP}$
will correspond to unacceptable $D^{0}-\bar{D}^{0}$ mixing.

\section{Direct CP violation in D decays}
For singly Cabibbo suppressed (SCS) D decays, the SM penguin contributions can be safely neglected, as they are highly suppressed by the CKM factor $V_{cb}^\ast V_{ub}$, the GIM suppression $m_b^2/m_W^2$ and the loop factors (see, for example \cite{kagan} and references therein). However the $Z^\prime$-induced FCNC is only loop suppressed, which may provide large enough CP violation effects to account for the LHCb measurement. The relevant $\Delta C=1$ effective Hamiltonian is given by
\begin{align}
 H_{eff}^{\Delta C=1}&=\frac{G_F}{\sqrt{2}}\left [ \sum_{p=d,s}
 \lambda_p(C_1 Q_1^p + C_2 Q_2^p) +  \sum_{i=3}^{6} {\tilde C}_i(\mu) {\tilde Q}_i(\mu)
 \right ]+H.c.~,
\end{align}
with $\lambda_p=V_{cp}^\star V_{up}$ are the CKM factors. $Q_{1}^p=(\bar{p}c)_{V-A})(\bar{u}p)_{V-A}$ and $Q_{2}^p=(\bar{p}_\alpha c_\beta)_{V-A})(\bar{u}_\beta p_\alpha)_{V-A}$ are the SM current-current operators where $\alpha$, $\beta$ are color indices. By integrating out the right-handed $Z^{\prime}$ field, one obtains
${\tilde Q}_{3,5}=(\bar{u}c)_{V+A}\sum_{q}(\bar{q}q)_{V\pm A}$ and ${\tilde Q}_{4,6}=(\bar{u}_\alpha c_\beta)_{V+A}\sum_{q}(\bar{q}_\beta q_\alpha)_{V\pm A}$ with $q=u,d,s$. For order-of-magnitude estimation for $D \to KK$, $\pi\pi$ decays, we use naive factorization with the Wilson coefficients at leading order. The magnitude of direct CP violation is determined by the ratio of new physics amplitude over the SM amplitude
\begin{align}
\frac{A_{NF}(D \to PP)}{A_{SM}(D\to PP)}=\frac{{\tilde C}_4+{\tilde C}_3/N_c+r_\chi({\tilde C}_6+{\tilde C}_5/N_c)}{\lambda_p (C_1+C_2/N_c)}
\end{align}
where $r_\chi=2m_K^2/m_c(m_s+m_q)=2m_\pi^2/m_c(m_u+m_d)$ in the SU(3) flavor limit with $m_q=(m_u+m_d)/2$. $P=K$,$\pi$ and $N_c=3$ in the naive factorization. The $Z^\prime$-induced Wilson coefficients at leading order can be obtained at the scale $\mu \simeq m_t$ as
\begin{align}
{\tilde C}_{4,6}=-3{\tilde C}_{3,5}=\frac{\alpha_s(m_t) g_X^2 \sin 2\theta e^{-i\delta}}{64\sqrt{2}\pi G_F m_{Z^\prime}^2} E_0(m_t^2/m_{Z^\prime}^2)
\end{align}
with the loop function \cite{UT-KOMABA-80-8}
\begin{align}
E_0(x)=&-\frac{2}{3}\ln x +\frac{x(18-11x-x^2)}{12(1-x)^3} 
       +\frac{x^2(15-16x+4x^2)}{6(1-x)^4}\ln x ~.
\end{align}
Notice that the renormalization group evolution of ${\tilde C}_i$ is the same as that of the SM QCD penguin operators with $L\leftrightarrow R$.
The Wilson coefficients at the scale $\mu_c$ can then be evaluated as
\begin{align}
{\tilde C}(\mu_c)=U_5(\mu_c, m_t){\tilde C}(m_t)
\end{align}
with the expression of $U_5$ given in \cite{hep-ph/9512380}. Here we have ignored the $b$ quark mass threshold for simplicity. As the U-spin symmetry predicts
$A_{CP}(K^+ K^-)=-A_{CP}(\pi^+ \pi^-)$, the LHCb evidence implies $A_{CP}(K^+ K^-)\simeq -0.0041 \pm 0.0012$ in the flavor symmetry limit. Numerically we take $m_c(m_c)=1.64$ GeV and $m_s=100$ MeV, $m_q=4.5$ MeV at $\mu=2$ GeV. Assuming the maximal CP phase $\delta=\pi/2$, we show in Fig. \ref{DCPV} the contour plot of $A_{CP}(K^+ K^-)$ as a function of the parameters $M_{Z^\prime}$ and $g_X \sqrt{\sin 2\theta}$.
\begin{figure}[htbp]
\begin{center}
\includegraphics[angle=0,width=8cm]{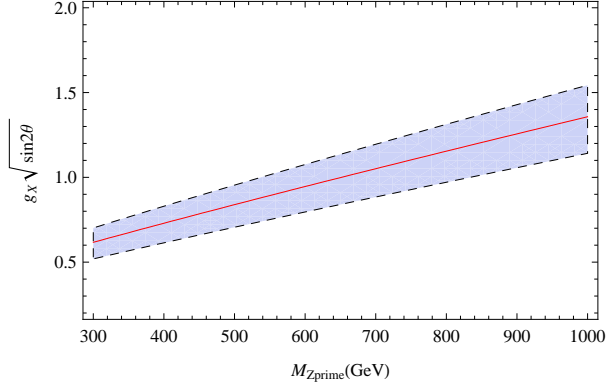}
\caption{Contour plot of the direct CP violation of $D \to K^+ K^-$ as a function of the parameters $M_{Z^\prime}$ and $g_X \sqrt{\sin 2\theta}$.  The solid red line represents the experimental central value and the light
blue (grey) region corresponds to one sigma contour.}
\label{DCPV}
\end{center}
\end{figure}

\section{Collider Implications}
As we discussed earlier, the up FCNC model is motivated to explain the
top quark forward-backward asymmetry. Figure \label{DCPV} shows
the best fit parameter region for such $Z^{\prime}$. The dominant
contribution for top quark $A_{FB}$ is through $ut$ coupling
which is $g_{X}\cos\theta$ in the above model. For a particular choice of $\theta\in\{0,\pi/4\}$, the
parameter space is consistent with the 1~$\sigma$ fitting of $A^{t\bar{t}}_{FB}(M_{t\bar{t}}>450~\text{GeV})$
\cite{jing}.

The flavor violating processes not only appear in the low energy physics but also appear in
the collider experiments. However, due to the challenge in the identification of the light quarks
states, only when the flavor violation involves top quarks directly, the measurements become
possible. In these models with up FCNC, both $\Delta F=1$ and $\Delta F=2$ effects
are observable at hadron collider. The $\Delta F=1$ processes correspond to single top
production associated with the $Z^{\prime}$ $g c \to t Z^{\prime}$ or $g u \to t Z^{\prime}$.
For Abelian model, $Z^{\prime}$ can also mediate $uu \to tt$ or $cc\to tt$.
The Large Hadron Collider~(LHC) is a proton-proton collider with center-of-mass energy
7~TeV in the first two years running. The same-sign positive top
quark pair ($uu\to tt$) becomes particular interesting at the LHC given its
large $u$-valence quark parton flux. However, the $\sigma_{uu\to tt}$ is proportional
to $\cos^{4}\theta$ while the $\sigma_{cc\to tt}$ has a factor as $\sin^{4}\theta$. The bounds from
flavor physics is only on $\sin 2\theta$.
In addition, without phase-space suppression at the 7 TeV LHC, the $t Z^{\prime}$ associate production is significant. Since $Z^{\prime}$ equally
decays into $u\bar{t}$ and $t\bar{u}$, the associated production $t Z^{\prime}$ or $\bar{t} Z^{\prime}$
will contribute to $tt+j$, $\bar{t}\bar{t}+j$  and $t\bar{t}+j$ final states.
And the $t\bar{t}+j$ will appear in the inclusive $t\bar{t}$ search.
However, it has been studied in \cite{jing}, the best-fit parameter space to explain the top quark $A_{FB}$
via  $Z^{\prime}$ is largely excluded by the Tevatron/LHC same-sign top search and the inclusive
$t\bar{t}$ at the LHC for $t\bar{t}+j$. We will not use the fitting parameter from top quark $A_{FB}$.
Instead, we focus on the $Z^{\prime}$ that can explain the $\Delta A_{CP}$ alone where the $ut$ coupling
does not dominate the $Z^{\prime}$ penguin. With larger $ct$ coupling, the direct search bound is weaker.
To illustrate the feature, we take the $\theta=\pi/4$ so that $\sin 2\theta$ reaches its maximal.
\begin{figure}[htbp]
\begin{center}
\includegraphics[angle=0,width=8cm]{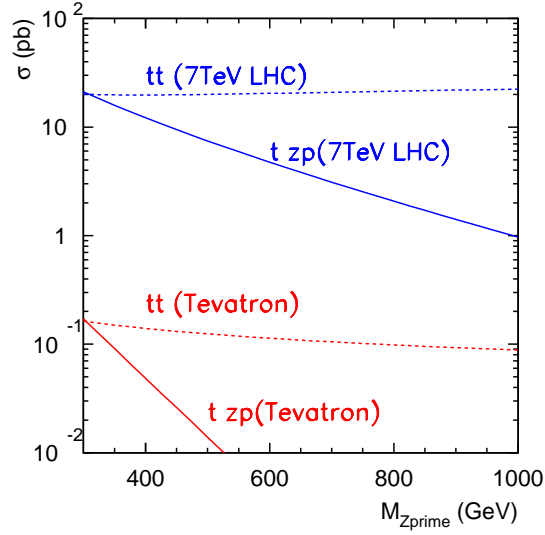}
\caption{Production cross sections for $\sigma(pp\to tt+\bar{t}\bar{t})$ at Tevatron and the 7 TeV LHC and the
 $\sigma(pp\to t Z^{\prime}+\bar{t}Z^{\prime}$) at Tevatron and the 7 TeV LHC. The coupling is taken to be $\theta=\pi/4$. }
\label{uutt}
\end{center}
\end{figure}

Figure \ref{uutt}  gives the $pp\to tt$ production rate at Tevatron and the 7 TeV LHC
with $\theta=\pi/4$ in the Fig. \ref{DCPV}.
The $p\bar{p}\to tt+\bar{t}\bar{t}$ at Tevatron is below 0.2~pb for these best fit points and
this corresponds to about 10 pure-leptonic same-sign top events with one b-tagging (50\% tagging efficiency) before any cut for integrated luminosity of 2~fb$^{-1}$.
CDF measured 3 events for 2~fb$^{-1}$ \cite{Aaltonen:2008hx} with the acceptance range from 1.5$\%$ to 3$\%$. It is still consistent with the measurements.  At LHC with 35~pb$^{-1}$,
the prediction is about 20 events of pure-leptonic same-sign top with one b-tagging.
Since $tt$ production is mostly $t$-channel, one expects the cut efficiency for the
forward-backward region top quarks are less. The latest LHC observation is 2 events
for  35~pb$^{-1}$ \cite{cms}. It seems to be still consistent with the observation at this moment.
However, given its large rate, the parameter region should soon be probed by the
CMS or ATLAS experiments. The $t Z^{\prime}$ production which comes into the inclusive $t\bar{t}$ search is
still within the error bar of the measurements.

\section{$D^{0}-\bar{D}^{0}$ mixing}

Any theory that contributes to $c\to u q\bar{q}$ is inevitable to
generate the $\Delta C=2$ process of $D^{0}-\bar{D}^{0}$ mixing. In the SM, $D^{0}-\bar{D}^{0}$ mixing is very slow due to the GIM mechanism, which is particularly effective in D meson since the $b$ quark contribution is accidentally suppressed by a very small CKM factor. But new physics without flavor suppression could easily saturate or even badly violate the experimental bound. The strongest bound comes from the Belle results \cite{arXiv:0704.1000} $x=(0.80 \pm 0.29 \pm 0.17)\%$, $y=(0.33 \pm 0.24 \pm 0.15)\%$, which leads to \cite{hep-ph/0703235}
\begin{align}\label{D-bound}
\vert M_{12}^D \vert \lesssim 1.2 \times 10^{-14} \mbox{GeV}
\end{align}
assuming CP conservation in mixing. Otherwise, the bound would be relaxed by a factor $\sim 2$.

It is straightforward to evaluate the $Z^\prime$ contribution to the $D^{0}-\bar{D}^{0}$ mixing,
\begin{align}
M_{12}^D=\frac{g_X^4 \sin^2 2\theta}{1536\pi^2 m_{Z^\prime}^2} \left ( \frac{\alpha_s(m_t)}{\alpha_s(\mu_c)}\right )^{6/23}
F(x_t,x_t) f_D^2 m_D
\end{align}
with $x_t=m_t^2/m_{Z^\prime}^2$.
The vacuum insertion approximation has been adopted in the above for simplicity
\begin{align}
\langle D^0 \vert (\bar{u}c)_{V+A} (\bar{u}c)_{V+A} \vert \bar{D}^0 \rangle =\frac{8}{3}m_D^2 f_D^2
\end{align}
and the Inami-Lim loop function F(x,y) \cite{UT-KOMABA-80-8} reads
\begin{align}
F(x,x)=\frac{4+4x-15x^2+x^3}{4(1-x)^2}+\frac{x(4-4x-3x^2)}{2(1-x)^3}\ln x
\end{align}
in the limit $y\rightarrow x$. Taking $f_D=220$ MeV, one finds unfortunately that the $Z^\prime$-induced $D^{0}-\bar{D}^{0}$ mixing
is two to three orders of magnitude larger than the experimental bound Eq. (\ref{D-bound}), for the favored parameter region shown in Fig. \ref{DCPV}.

\section{Conclusion}
In this paper, we investigate the up FCNC models of flavor violation gauge interaction in the up sector to explain the LHCb recent observed large $\Delta A_{CP}$ ($A_{CP}(D^{0}\to K^{+}K^{-})-A_{CP}(D^{0}\to \pi^{+}\pi^{-})$).  To illustrate the feature, we study an Abelian model that only right-handed up quarks is charged under it and the $1-3$ coupling is maximized. The simultaneous $1-3$ $2-3$ mixing is realized by a quark mixing of $1-2$ generation.  Given the easy identification of top quark, the model can be directly tested by $\Delta F=1$ and $\Delta F=2$ processes at the hadron colliders as associated top production $g c \to t Z^{\prime}$ or same-sign top scattering $u u\to t t$. The direct search bounds are still consistent with the assumption that $ut$ and $ct$ couplings are equal but the same-sign top scattering bound is expected to be reached very soon.

However, since there is no CKM-like suppression, the corresponding parameter space for generating $\Delta A_{CP}$ is completely excluded by the $D^{0}-\bar{D}^{0}$ mixing. We conclude that the up FCNC type models cannot explain the $\Delta A_{CP}$ while to be consistent with the $D^{0}-\bar{D}^{0}$ mixing constraint. On the other hand,  a model as SM with fourth family extension has better chance to explain the large $\Delta A_{CP}$ consistently.

\section*{Acknowledgement}
We would like to thank Ming-xing Luo for useful discussion.
%ML is also supported in part by the National Science Foundation of China (10875103), National Basic Research Program of China (2010CB833000), and Zhejiang University Group Funding (2009QNA3015).
KW is supported by the Fundamental Research Funds for the Central Universities (2011QNA3017).
GZ is supported by the National Science Foundation of China (No. 11075139 and No.10705024) and the Fundamental Research Funds for the Central Universities.

\end{document}